\documentclass[twocolumn]{aastex63}
\usepackage[normalem]{ulem}
\usepackage{bm}

\received{2020 Nov 8}
\revised{2021 Jan 22}
\accepted{2021 Feb 6}
\submitjournal{PSJ}

\shorttitle{The Reflectance of Cold Classical TNOs}
\shortauthors{Seccull, Fraser, \& Puzia}

\begin{document}

\title{The Reflectance of Cold Classical Trans-Neptunian Objects in the Nearest Infrared}

\correspondingauthor{Tom Seccull}
\email{tom.seccull@noirlab.edu}

\author[0000-0001-5605-1702]{Tom Seccull}
\affiliation{Gemini Observatory/NSF's NOIRLab, 670 N. A'ohoku Place, Hilo, HI 96720, USA}
\affiliation{Astrophysics Research Centre, Queen's University Belfast, University Road, Belfast, BT7 1NN, UK}

\author[0000-0001-6680-6558]{Wesley C. Fraser}
\affiliation{Herzberg Institute of Astrophysics, 5071 West Saanich Road, Victoria, BC V9E 2E7, Canada}

\author[0000-0003-0350-7061]{Thomas H. Puzia}
\affiliation{Institute of Astrophysics, Pontificia Universidad Cat\'olica de Chile, Av. Vicu\~na MacKenna 4860, 7820436, Santiago, Chile}



\begin{abstract}
Recent photometric surveys of Trans-Neptunian Objects (TNOs) have revealed that the cold classical TNOs have distinct z-band color characteristics, and occupy their own distinct surface class. This suggested the presence of an absorption band in the reflectance spectra of cold classicals at $\lambda>0.8~\mu$m. Here we present reflectance spectra spanning $0.55-1.0~\mu$m for six TNOs occupying dynamically cold orbits at $a\sim44$~au. Five of our spectra show a clear and broadly consistent reduction in spectral gradient above $0.8~\mu$m that diverges from their linear red optical continuum and agrees with their reported photometric colour data. Despite predictions, we find no evidence that the spectral flattening is caused by an absorption band centered near $1.0~\mu$m. We predict that the overall consistent shape of these five spectra is related to the presence of similar refractory organics on each of their surfaces, and/or their similar physical surface properties such as porosity or grain size distribution. The observed consistency of the reflectance spectra of these five targets aligns with predictions that the cold classicals share a common history in terms of formation and surface evolution. Our sixth target, which has been ambiguously classified as either a hot or cold classical at various points in the past, has a spectrum which remains nearly linear across the full range observed. This suggests that this TNO is a hot classical interloper in the cold classical dynamical range, and supports the idea that other such interlopers may be identifiable by their linear reflectance spectra in the range $0.8-1.0~\mu$m.
\end{abstract}

\keywords{Classical Kuiper belt objects, Trans-Neptunian Objects, spectroscopy}


\section{Introduction} \label{sec:intro}
The cold classical Trans-Neptunian Objects (TNOs) are minor planets on non-resonant, non-scattering, low inclination ($i\lesssim5^{\circ}$), and low eccentricity ($e<0.24$) heliocentric orbits with semimajor axes between 40 and 50~au \citep{2008ssbn.book...43G,2011AJ....142..131P,2018ApJS..236...18B}. The cold classicals are distinct from the dynamically excited (hot) TNOs in a number of ways, including their lack of dynamical excitation; their higher fraction of surviving, and in particular widely separated, binary pairs \citep{2006AJ....131.1142S,2008Icar..194..758N,2010ApJ...722L.204P}; their distinct size distribution and its lack of large objects \citep[$Diameter\gtrsim400$~km; e.g.][]{2014ApJ...782..100F,2020tnss.book..153M}; and their unimodally (rather than bimodally) distributed colors and albedos, which on average respectively sit at the very red and high ends of the color and albedo distributions of TNOs as a whole \citep{2003ApJ...599L..49T,2009Icar..201..284B,2012ApJ...749...33F,2014ApJ...793L...2L,2014A&A...564A..35V,2017AJ....154..101P,2018PASJ...70S..40T,2019ApJS..243...12S}. Notably the cold classical TNOs are typically not quite as red as the very reddest TNOs such as 225088 Gonggong \citep[2007~OR$_{10}$;][]{2011ApJ...738L..26B}. Taken together the distinct properties of the cold classicals support the prediction that they formed and evolved in-situ, separately from the other TNO populations, and that they emerged mostly unscathed from the period of giant planet migration thought to have emplaced the dynamically hot TNOs into the modern trans-Neptunian Belt \citep{2008Icar..196..258L,2010ApJ...722L.204P,2011ApJ...738...13B,2012ApJ...746..171W,2015AJ....150...68N}.

After setting aside any interlopers sharing their orbital parameter space \citep[such as the blue binaries;][]{2017NatAs...1E..88F}, the cold classical TNOs consitute the best preserved samples of material derived from the outermost reaches of the Solar protoplanetary disk at the time of planet formation. They have been spared significant surface evolution from most of the common sources that affect minor planets. The delivery of collisional and accretional heat to the cold classicals is limited by the relatively low frequency and velocities of the collisions they experience \citep{2012MNRAS.425.1492D,2014ApJ...782..100F,2019ApJ...872L...5G,2020Sci...367.3999S}. TNOs as small as most cold classicals ($D<400$~km) tend to have low densities \citep[e.g.][]{2007Icar..191..286G,2012A&A...541A..93M,2012A&A...541A..92S,2012A&A...541A..94V,2014A&A...564A..35V,2013A&A...557A..60L,2017A&A...608A..19K}, suggesting that they may have a low rock/ice ratio, and consequently a lower capacity for internal radiogenic heating \citep[e.g.][]{2008ssbn.book..243C}. At large heliocentric distances of $40-50$~au weak insolation doesn't drive significant thermal evolution of planetary surfaces \citep[e.g.][]{2011A&A...529A..71G}. By residing at these distances since their formation \citep{2010ApJ...722L.204P,2020tnss.book...25M}, the cold classicals are also predicted to have received relatively low doses of photonic and ionic radiation from both the Sun and galactic sources \citep{2003EM&P...92..261C,2006AIPC..858..372C,2008ssbn.book..507H}. 

Due to the relatively primitive nature of the materials on their surfaces, the cold classical TNOs are enticing targets for spectroscopic study, and may provide one of the most direct observational routes to measurement of the chemical and thermal conditions that prevailed in the outer protoplanetary disk of the young Sun. Due to their faintness, however, they also prove to be challenging observational targets. To date only four TNOs that have been solidly classified as cold classicals have been observed spectroscopically (see Table \ref{tab:prev}).

\begin{table*}
\begin{centering}
\caption{Reflectance Spectra of Cold Classical TNOs Reported to Date}
\label{tab:prev}
\begin{tabular}{
p{5cm}                      p{5.cm}             p{7cm}
}
\hline\hline
TNO                                    & Wavelength Ranges Covered, $\mu$m          & References  \\[1pt]
\hline
58534 Logos (1997~CQ$_{29}$)            & $0.6-0.88$                    & \citep{2001AA...378..653B} \\[1pt]
66652 Borasisi-Pabu (1999~RZ$_{253}$)  & $1.4-2.4$                     & \citep{2008AJ....135...55B} \\[1pt]
79360~Sila-Nunam (1997~CS$_{29}$)       & $0.4-0.88$; $1.4-2.5$  & \citep{2001AA...378..653B,2005AJ....130.1299G}\\[1pt]
486958~Arrokoth (2014~MU$_{69}$)        & $1.2-2.5$                     & \citep{Sterneaaw9771,2020Sci...367.3705G} \\[1pt]
\hline
\end{tabular}\\[2pt]
\end{centering}
\end{table*}

Within the signal-to-noise (S/N) of the observational data and the wavelength ranges covered so far, the reflectance spectra of cold classicals are almost completely featureless. In the optical range they show only continuum with a strong positive gradient, while in the near-infrared (NIR) they appear neutrally reflective or very slighly blue \citep{2005AJ....130.1299G,2020Sci...367.3705G,2008AJ....135...55B}. The only absorption bands confidently identified in the reflectance spectrum of a cold classical were observed at $2.271~\mu$m and $2.338~\mu$m in the NIR spectrum of Arrokoth by the NASA \textit{New Horizons} probe; these bands were attributed to surface methanol ice \citep{2020Sci...367.3705G}. The red optical color of the cold classicals is typically attributed to the presence of complex macromolecular organic residues on their surfaces \citep[e.g.][]{2015Icar..252..311D}. Such materials are predicted to be primarily comprised of a highly diverse agglomeration of complex hydrocarbon molecules, and are known to be produced through radiolytic and/or photolytic processing of the simple molecular ices from which the TNOs are thought to have formed \citep[e.g.][]{1989Icar...79..350K,2005AdSpR..36..178C,2006ApJ...644..646B,2014ApJ...788..111M,2015ApJ...812..150M}. The featureless nature of the optical and NIR reflectance spectra of TNOs has so far precluded robust quantitative chemical analysis of their surfaces, however.

Recently the reflectance properties of the cold classicals were reported to be distinct from those of the dynamically hot TNOs based on photometric colors obtained by both the Colors of the Outer Solar System Origins Survey \citep[Col-OSSOS;][]{2017AJ....154..101P} and an independent study conducted by \citet{2018PASJ...70S..40T}. From the colors measured by both teams, the dynamically hot TNOs were inferred to have average reflectance spectra with only a small decrease in gradient across the range $0.6-1.0~\mu$m, if any at all. By constrast, cold classicals were found to occupy a unique space in $g-r$/$r-z$ color space, implying a reduced spectral gradient at $\lambda\sim0.9~\mu$m compared to that across the same wavelength region for the dynamically hot TNOs with similar optical color \citep{2017AJ....154..101P,2018PASJ...70S..40T}. 

The distinct spectroscopic behaviour of the cold classicals may set them apart from dynamically hot TNOs with similar optical redness in terms of composition; a distinction like this warrants spectroscopic investigation. With the broad aim of spectroscopically confirming and characterising the distinct spectral behaviour of the cold classical TNOs at $\lambda>0.8~\mu$m, we set out to obtain a sample of their reflectance spectra.

\section{Target Selection}
Our original sample was comprised of nine targets. 505476 (2013~UL$_{15}$) and 511552~(2014~UE$_{225}$) were selected because they had published colours obtained by the Col-OSSOS survey, and were predicted by \citet{2017AJ....154..101P} to have a large divergence from linearity in their reflectance spectra at $\lambda>0.8~\mu$m. The other seven TNOs were selected from the catalog of \citet{2015A&A...577A..35P} if they were dynamically classified as a cold classical, had a full set of $BVRI$ photometric colours, and were bright enough ($R<22.5$) that spectroscopic observations would be obtainable, reducable, and useful for characterisation and analysis. Of the nine initially selected targets, six were successfully observed.

\section{Observations}

\begin{table*}
\begin{center}
\caption{Observation Log}
\label{tab:obslog}
\begin{tabular}{p{3.cm}p{4.4cm}p{1.cm}p{0.5cm}p{1.4cm}p{1.4cm}p{0.8cm}p{0.7cm}p{0.6cm}}
\hline\hline
Target\textsuperscript{\textit{a}} & UT Observation Date $\vert$ Time & T\textsubscript{exp}, s & N\textsubscript{exp} & Airmass & IQ, \arcsec & $\Delta$, au & $r$, au & $\alpha$, $^{\circ}$ \\[1pt]
\hline
HD~225194 & 2019-08-29 $\vert$ 05:25:06--05:33:21 & 0.3 & 2 & 1.29 & 0.81-1.11  \\[1pt]
511552 (2014~UE$_{225}$) & 2019-08-29 $\vert$ 05:33:39--09:16:49 & 500.0 & 16 & 1.28--1.54 & 0.69--1.10 & 46.02 & 46.59 & 1.03\\[1pt]
\textbf{HD~8291} & 2019-08-29 $\vert$ 09:17:03--09:26:16 & 0.3 & 2 & 1.28 & 0.96--1.03 \\[1pt]
\hline
\textbf{HD~209562} & 2019-08-30 $\vert$ 02:21:13--02:30:51 & 0.3 & 2 & 1.10 & 0.81--1.09 \\[1pt]
275809 (2001~QY$_{297}$) & 2019-08-30 $\vert$ 02:31:53--05:18:54 & 500.0 & 12 & 1.02--1.06 & 0.78--1.43 & 42.94 & 43.94 & 0.13\\[1pt]
HD~218251 & 2019-08-30 $\vert$ 05:19:07--05:30:20 & 0.4 & 2 & 1.02 & 0.85--0.96 \\[1pt]
\hline
BD-00~4557 & 2019-08-30 $\vert$ 05:40:27--05:48:51 & 0.7 & 2 & 1.11 & 0.80--0.83 \\[1pt]
138537 (2000~OK$_{67}$) & 2019-08-30 $\vert$ 05:49:04--08:24:16 & 500.0 & 12 & 1.09--1.32 & 0.66--1.02 & 39.12 & 40.05 & 0.56\\[1pt]
\textbf{HD~224448} & 2019-08-30 $\vert$ 08:24:42--08:31:55 & 0.4 & 2 & 1.36 & 1.01--1.16 \\[1pt]
\hline
\textbf{HD~142331} & 2019-08-30 $\vert$ 23:27:51--23:36:52 & 0.3 & 2 & 1.13 & 0.72--1.01 \\[1pt]
2001~HZ$_{58}$ & 2019-08-30 $\vert$ 23:43:53--01:33:48 & 500.0 & 8 & 1.19--1.61 & 0.71--1.61 & 43.89 & 43.75 & 1.31\\[1pt]
HD~147284 & 2019-08-31 $\vert$ 01:42:11--01:50:03 & 0.4 & 2 & 1.42 & 0.71--0.81 \\[1pt]
\hline
66652 Borasisi-Pabu & 2019-08-31 $\vert$ 01:59:50--04:18:28 & 500.0 & 12 & 1.16--1.78 & 0.70--0.90 & 41.27 & 42.23 & 0.41\\[1pt]
\textbf{HD~224448} & 2019-08-31 $\vert$ 04:18:47--04:26:34 & 0.4 & 2 & 1.21 & 0.78--0.81 \\[1pt]
\hline
\textbf{HD~8291} & 2019-08-31 $\vert$ 04:26:51--04:35:27 & 0.3 & 2 & 1.58 & 0.81--0.81 \\[1pt]
505476 (2013~UL$_{15}$) & 2019-08-31 $\vert$ 04:35:38--09:43:25 & 500.0 & 24 & 1.18--1.5 & 0.46--0.93 & 42.79 & 43.50 & 0.96\\[1pt]
HD~2141 & 2019-08-31 $\vert$ 09:43:30--09:49:51 & 0.4 & 2 & 1.47 & 0.56--0.67 \\[1pt]
\hline
\end{tabular}\\[2pt]
\end{center}
\small{\textbf{Notes:} For each target we present the UT observation date and time, the integration time per exposure (T\textsubscript{exp}), the number of exposures (N\textsubscript{exp}), and the airmass at which they were observed. The range of estimated image quality (IQ) values presented for each target is the range of Full Widths at Half Maximum measured from Moffat profiles \citep{1969A&A.....3..455M} fitted to the set of spatial profiles produced by median collapsing each of a target's reduced 2D spectroscopic exposures along the dispersion axis. For the TNOs we present their geocentric distances ($\Delta$), heliocentric distances ($r$), and phase angles ($\alpha$) at the time of observation. The apparent magnitude of our targets lay in the range $21.5\lesssim{r}\lesssim23$. \\
\textsuperscript{\textit{a}} Stars presented in bold are those ultimately used to calibrate the spectra of their associated TNOs.}
\end{table*}

Each spectrum presented here was observed during a visitor mode observing run at the European Southern Observatory's (ESO's) Very Large Telescope (VLT) that took place during the three nights between 2019 Aug 28 and 31. Table \ref{tab:obslog} presents a log of our successful observations. At the VLT we used the FOcal Reducer/low dispersion Spectrograph 2 \citep[FORS2;][]{1998Msngr..94....1A} mounted on the 8.2~m UT1(Antu) unit telescope. FORS2 is a multipurpose optical spectrograph, imager, and polarimeter, which we used in Long-Slit Spectroscopy mode to observe our TNOs and Solar calibrators. FORS2 was configured with the red sensitive MIT CCD detector, the standard resolution collimator, and a longslit of width 1.6\arcsec. The non-standard \verb.GRIS_200I+28. grism was installed in FORS2 for our observations, offering a resolving power of $\lambda/\Delta\lambda = 380$~ at $0.745~\mu$m and a wavelength coverage of $0.55-1.1~\mu$m. The extremely low S/N of the spectra at the longest wavelengths would effectively limit the extent of our useful data to the range $0.55-1.0~\mu$m, however. The \verb.GRIS_200I+28. grism has an OG550 order separation filter cemented to it, so no extra order blocking filter was required.

For all TNOs and Solar calibrators we used a two-point repeating spatial dither pattern between spectroscopic exposures centered at $+1\arcsec$ and $-1\arcsec$ from the center of the spatial axis of the slit. To mitigate the effects of atmospheric differential refraction \citep{1982PASP...94..715F} we aligned the slit to the parallactic angle before commencing spectroscopic observations of each target, and for each TNO the slit was realigned to the parallactic angle after every four exposures (approximately once every 45 minutes). All exposures were read out using the standard 100kHz,2x2,high readout mode.

We ensured that the observations of each TNO were bracketed by observations of a Solar calibrator obtained at a similar airmass and pointing. The spectra of these calibrators were used to cancel the Solar spectrum from those of the TNOs and derive the TNO reflectance spectra; they also double as telluric standards. Only one bracketing calibrator spectrum was required, but observation of two provided redundancy in the event of unexpected changes in conditions. At least one Solar calibrator for each TNO was selected from published catalogs of Solar twins and analogs \citep{2009A&A...508L..17R, 2014A&A...572A..48R}. If the second calibrator was not a Solar twin or analog it was a star with both a reported spectral type and $B-V$ color that were close to Solar \citep{2000A&AS..143....9W}. Ultimately we were able to calibrate each TNO spectrum with an associated Solar twin or Solar analog spectrum, and we therefore expect the gradients and shapes of the TNO reflectance spectra to be accurate. All our spectra were observed under clear (sky 90\% cloudless at elevation~$>30^{\circ}$, transmission variation~$<10\%$) or photometric (sky cloudless, transmission variation~$<2\%$) conditions as defined by ESO. We therefore expect any effects of cloud cover on the continuum colour of our spectra to be minimal. Conditions also remained consistently dry thoughout our run (Precipitable Water Vapor; PWV~$<1.5$~mm).  

\section{Data Reduction}
Standard spectroscopic data reduction steps including bias subtraction, flat-fielding, and wavelength calibration were conducted for all our raw specta with the ESO FORS data reduction pipeline (v.~5.4.3\footnote{ftp://ftp.eso.org/pub/dfs/pipelines/instruments/fors/fors-pipeline-manual-5.10.pdf}) in the ESO Reflex data processing environment \citep[v.~2.9.1;][]{2013A&A...559A..96F}. The spectra were not corrected for instrument response, as the response of FORS2 is stable over a period of several days, and any response calibration applied to a TNO spectrum and a Solar calibrator star spectrum would only be cancelled out once the former was divided by the latter later on.

While not a problem for the 2D spectra of the bright Solar calibrator stars, low level fringing was identified at longer wavelengths in the 2D spectra of the TNOs, so a fringe frame was constructed for each TNO to remove them. First, for each 2D spectrum associated with a given TNO, a linear profile was fitted and subtracted from each pixel column parallel to the spatial axis to approximately cancel the illumination pattern of sky emission lines. Next, each spatial pixel column within each 2D spectrum was sigma-clipped at $\pm2\sigma$ before all the sigma-clipped 2D spectra were mask-median combined to form a fringe frame that didn't contain spectra of any sources. This fringe frame was then subtracted from each 2D spectrum of the TNO to cancel the fringes. A linear profile was then fitted to, and subtracted from, each pixel column parallel to the spatial axis again to remove the sky emission line illumination pattern. This procedure makes the simple assumption that the sky line illumination has a spatially linear profile, but tests making use of a second order polynomial didn't show significant improvement in the illumination pattern cancellation.

Sky subtraction, cosmic-ray removal, and extraction of the spectra were performed using a method similar to that used by \citet{2018ApJ...855L..26S}. Briefly, Moffat functions \citep{1969A&A.....3..455M} were fitted to the spatial profile of the 2D wavelength-calibrated spectrum at many locations along the dispersion axis in order to track the wavelength dependent center and width of the spectrum's spatial profile. Sky region boundaries were defined at $\pm3$ Full-Widths at Half Maximum (FWHM) from each Moffat profile center with sky outside these boundaries. 15 iterations of sigma clipping at 5$\sigma$ were separately conducted in the sky and target regions to remove cosmic rays. In each unbinned wavelength element the median background was subtracted. While the TNO spectra had effectively been background subtracted during fringe removal, performing a simple median sky subtraction was useful in making small improvements to the quality of the background removal. 

Another round of Moffat fitting was conducted for the sky-subtracted spectrum and extraction limits were defined at $\pm2.5$ FWHM from each of the Moffat profile centers. During this process, pixels that formerly contained cosmic rays had their values replaced using a process that is similar to that used in the ESO X-Shooter pipeline for the same purpose \citep{2010SPIE.7737E..28M}; details are presented in the X-Shooter pipeline manual\footnote{see the Standard Exraction algorithm; ftp://ftp.eso.org/pub/dfs/pipelines/instruments/xshooter/xshoo-pipeline-manual-3.3.5.pdf}. Within the defined extraction limits the flux was summed for each wavelength element to form the 1D spectrum.

The spectra were all then corrected for atmospheric extinction using $f_{C}(\lambda) = f(\lambda)10^{0.4ak(\lambda)}$, where $f_{C}(\lambda)$ and $f(\lambda)$ are respectively the extinction corrected and uncorrected spectra, $a$ is the median airmass at which the spectrum was observed, and $k(\lambda)$ is the FORS2 extinction coefficient (taken from the FORS2 data reduction pipeline) interpolated to the resolution of the spectra.

Once corrected for atmospheric extinction, the 1D spectra for each target were median stacked. For Borasisi-Pabu, 2000~OK$_{67}$, 2001~QY$_{297}$, and 2001~HZ$_{58}$, however, 1, 3, 3, and 2 spectra were respectively excluded from their final stacks due to their very low S/N or large sky emission line  residuals. The stacked spectra of all other targets, including the solar calibrators, were produced using all of their available spectroscopic exposures. Solar calibration of the stacked TNO spectra was performed by dividing them by their associated stacked Solar calibrator spectrum, resulting in the production of the reflectance spectrum of each TNO. The reflectance spectra were then binned using the bootstrapping method described by \citet{2019AJ....157...88S} to estimate their uncertainties and boost their S/N at the expense of wavelength resolution. Binning factors for each spectrum are presented in Fig. \ref{fig:unbin}.

\section{Results \& Analysis}\label{sec:res}

\begin{figure*}
\centering
\includegraphics[scale=0.5]{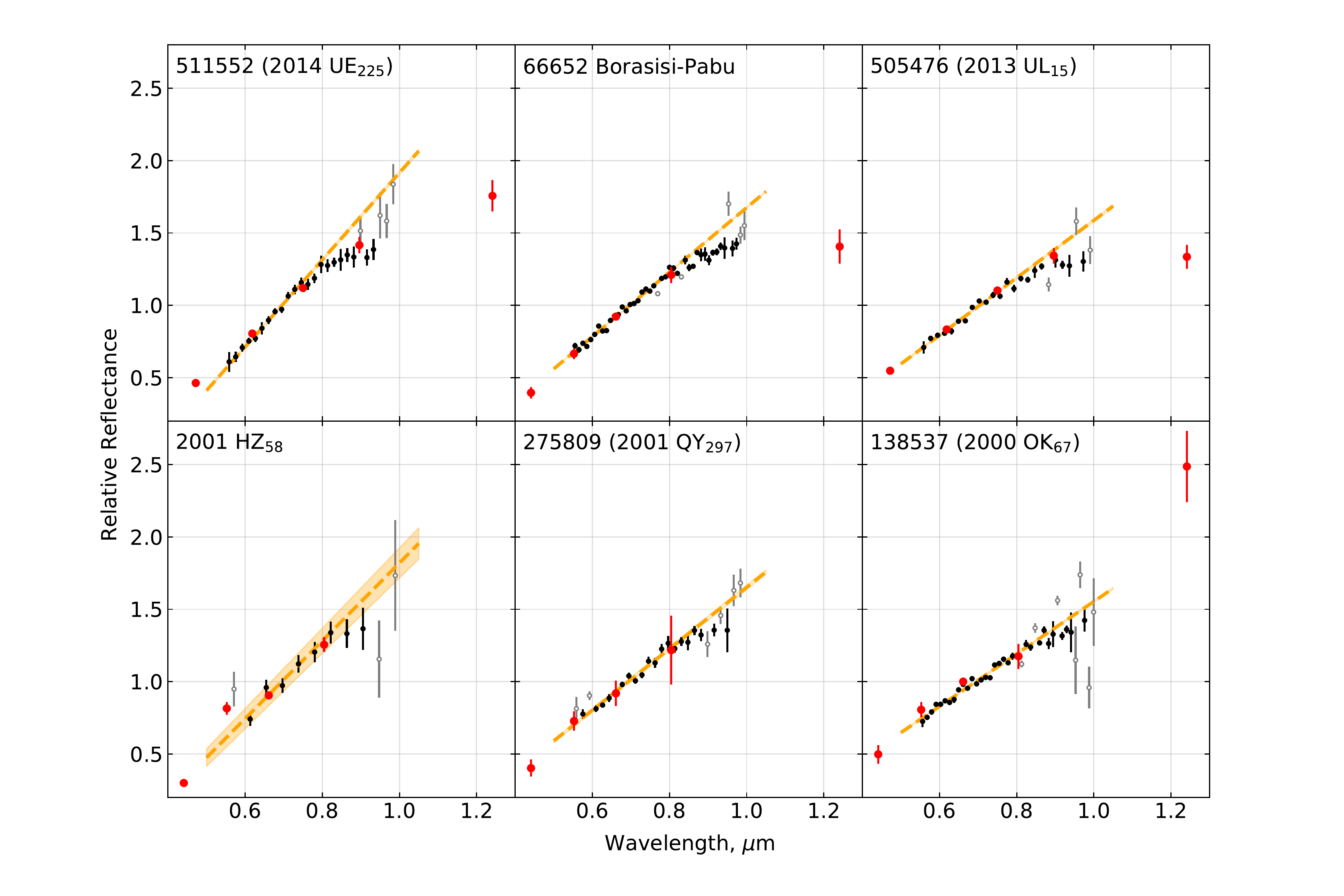}
\caption{Reflectance spectra of six dynamically cold TNOs at $40<r<47$~au. Our observed spectra are plotted in black and cover the range $0.55-1.0~\mu$m. Grey points are significantly affected by residual sky lines. Orange dashed lines show a linear fit to the spectrum in the range $0.55-0.75~\mu$m extrapolated across the full wavelength coverage of the spectra, while the associated shaded regions denote the standard error of the linear fit. Red points are coarse reflectance spectra derived from published photometric colors \citep{2003Icar..161..501M,2004A&A...417.1145D,2015A&A...577A..35P,2017AJ....154..101P,2019ApJS..243...12S}. We use the following updated Col-OSSOS colors to derive the coarse reflectance spectra presented here: \textbf{2013~UL$_{15}$} $g-r=0.895\pm0.031$, $r-i=0.414\pm0.024$, $r-z=0.658\pm0.031$, $r-J=1.482\pm0.067$; \textbf{2014~UE$_{225}$} $g-r=1.04\pm0.017$, $r-i=0.468\pm0.024$, $r-z=0.753\pm0.031$, $r-J=1.817\pm0.067$. All datasets are scaled to unit reflectance at $0.7~\mu$m.  \label{fig:spec}}
\end{figure*}

\begin{table*}
\begin{center}
\caption{Cold Classical TNO Properties}
\label{tab:cctnoprop}
\begin{tabular}{
cccccccc
}
\hline\hline
TNO                        & $S'(0.65)$, $\%/(0.1~{\mu}m)$         & $\lambda_T$, ${\mu}m$     & $a$, au     & $e$   & $i$, $^{\circ}$ & $q$, au & $H_{V}$  \\[1pt]
\hline
66652 Borasisi-Pabu**      & $24.76\pm0.08$     & $>0.81$         & 43.98 & 0.092 & 0.563        & 39.95 & $6.12\pm0.07$          \\[1pt]
138537 (2000~OK$_{67}$)    & $19.76\pm0.15$     & $>0.87$         & 46.77 & 0.143 & 4.877        & 40.04 & $6.47\pm0.13$          \\[1pt]
275809 (2001~QY$_{297}$)** & $22.72\pm0.31$     & $>0.87$         & 44.17 & 0.085 & 1.547        & 40.42 & $5.86\pm0.31$         \\[1pt]
505476 (2013~UL$_{15}$)    & $22.28\pm0.20$     & $>0.80$         & 45.90 & 0.101 & 2.025        & 41.26 & $7.04\pm0.07$ \\[1pt]
511552 (2014~UE$_{225}$)   & $33.87\pm0.27$     & $>0.75$         & 43.73 & 0.066 & 4.504        & 40.84 & $6.54\pm0.04$ \\[1pt]
2001~HZ$_{58}$             & $29.12\pm0.91$     & $>0.86$         & 42.94 & 0.032 & 2.933        & 41.55 & $6.63^{+0.08}_{-0.29}$ \\[1pt]
\hline
Average                    & $25.41\pm2.12$     & $>0.83$        \\[1pt]
Average w/o 2000~OK$_{67}$ & $26.55\pm2.19$     & $>0.82$        \\[1pt]
\hline
\end{tabular}\\[2pt]
\end{center}
\small{\textbf{Notes:} For each TNO that we observed we present the properties that we compared in a search for evidence of any trends between their reflectance properties, their absolute $V$-band magnitudes ($H_{V}$), and their orbital semimajor axes, eccentricities, inclinations, and perihelia ($a$, $e$, $i$, and $q$, respectively). The gradients measured from our spectra across $0.55-0.75~\mu$m  ($S'(0.65)$) are shown alongside the minimum wavelength at which each spectrum diverges from a linear gradient ($\lambda_T$). Averages of these values and the standard errors of these averages are also presented for the full sample, and the full sample excluding the potential interloper 2000~OK$_{67}$. Methods used to estimate the $H_{V}$ values for our targets are described in section \ref{sec:notrends}. TNOs with marked with {*}{*} are known binaries.}
\end{table*}

Figure \ref{fig:spec} presents the reflectance spectra of the six TNOs we observed. These are the first optical reflectance spectra reported for any of these targets, and five of them \citep[excluding Borasisi-Pabu which has a NIR spectrum;][]{2008AJ....135...55B} are the very first to be reported for these objects at any wavelength. These TNOs have red and nearly linear spectra at $0.55-0.75~\mu$m, and all with the exception of 2000~OK$_{67}$ (see Section \ref{sec:interlopers}) have a clear transition in their reflectance spectra at $\lambda>0.75~\mu$m where their spectral gradients decrease towards the NIR. It is worth noting that a moderately strong telluric water absorption band exists in the range $0.9-1.0~\mu$m \citep[cf.][]{2015A&A...576A..77S}. While differences between the strength of this telluric band in the raw spectra of the TNOs and their solar calibrator stars have the potential to affect the shape of the reflectance spectra at $\lambda>0.9~\mu$m, the very stable dry conditions (PWV~$<1.5$~mm) under which our spectra were observed give us confidence that any curvature at these wavelengths is intrinsic to the TNOs themselves. We note that the $0.9-1.0~\mu$m telluric band cannot account for any curvature in the spectra at $\lambda<0.9~\mu$m. Our observation of this curvature spectroscopically supports inferences of the existence of such a flattening in the reflectance spectra of cold classicals from photometric color measurements \citep{2004A&A...417.1145D,2007AJ....134.2186D,2012ApJ...749...33F,2017AJ....154..101P,2018PASJ...70S..40T,2019ApJS..243...12S}.

Ancillary plots showing the appearance of our spectra at their native wavelength resolution (Fig. \ref{fig:unbin}), the consistent shape of the spectra of our Solar calibrators (Fig. \ref{fig:stars}), and the wavelength dependent S/N of our binned spectra (Fig. \ref{fig:snr}) can be found in the Appendix.

\subsection{Comparison to Published Photometry}

Since no optical reflectance spectra have yet been reported in the literature for any of our targets we compared our dataset to coarse reflectance spectra produced from published photometric colours using methods described by \citet{2002A&A...389..641H}. For Borasisi-Pabu, 2001~QY$_{297}$, and 2001~HZ$_{58}$, $BVRI$ reflectance points were determined using optical colors reported by \citet{2015A&A...577A..35P}. The simultaneous $V-J$ color measurement of Borasisi-Pabu reported by \citet{2003Icar..161..501M} was used to estimate its $J$-band reflectance. The $BVRIJ$ reflectance points for 2000~OK$_{67}$ were determined using simultaneously observed $BVRIJ$ colors reported by \citet{2004A&A...417.1145D}. $grizJ$ reflectance points for 2013~UL$_{15}$ and 2014~UE$_{225}$ were derived from colors published by the Outer Solar System Origins Survey \citep{2017AJ....154..101P,2019ApJS..243...12S}. We make use of updated values however, which reflect post-publication improvements to the photometry pipelines developed by that group (see Figure \ref{fig:spec}; Fraser et al., in prep.). Solar $BVRI$, $V-J$, $r-J$, and $griz$ photometric colors were respectively taken from \citet{2012ApJ...752....5R}, \citet{2012ApJ...761...16C}, \citet{2019ApJS..243...12S}, and the web pages of the Sloan Digital Sky Survey\footnote{\url{https://www.sdss.org/dr12/algorithms/ugrizvegasun/}}. Overall the photometrically derived reflectances of our targets are in good agreement with our reflectance spectra. We note that our reflectance spectra do not extend to wavelengths that are short enough to test for the presence of the non-linearity suggested to be present in the reflectance spectrum of 2001~HZ$_{58}$ by its $V$-band photometric point \citep{2015A&A...577A..35P}. In addition, some difference between our spectrum and the photometrically derived reflectance of 2000~OK$_{67}$ may be expected, as this object has been reported to have variable optical colors \citep{2002AJ....124.2279D}. 

\subsection{Measurements and a Search for Trends}\label{sec:notrends}
We measured the optical gradient of each reflectance spectrum across the wavelength range $0.55-0.75~\mu$m relative to the reflectance at $0.65~\mu$m using the bootstrapping method of \citet{2019AJ....157...88S}. Measured gradients are presented in Table~\ref{tab:cctnoprop} along with the average values of these measurements for this sample. The average optical gradient measured from our spectra (see Table~\ref{tab:cctnoprop}) is consistent with those previously determined from photometric colour measurements of cold classical TNOs \citep[e.g.][]{2012A&A...546A.115H}, and also gradients measured from the limited number of previously published optical reflectance spectra of cold classicals \citep[see][]{2001AA...378..653B}. We did not attempt to constrain the surface composition of our targets via use of \citet{hapke_2012} modelling on our reflectance spectra as they are featureless, and such analysis would therefore only return highly degenerate results.

To estimate the central wavelength of the transition in our spectra ($\lambda_{T}$), and constrain the gradient of the NIR continuum, we fitted a dual sloped spectral model to each spectrum, including J-band spectral photometry where available. The precision of this method was ultimately limited by our lack of spectral coverage in $J$ band, however, resulting in poor constraint of $\lambda_{T}$ and the NIR continuum gradient, making them insufficient for use in further analysis. A crude estimate of the lower limit of $\lambda_{T}$ was made for each spectrum by finding the lowest wavelength at which all redward datapoints have reflectances below the fitted linear continuum slope extrapolated redward from the optical region. These lower limits on $\lambda_{T}$ are presented in Table~\ref{tab:cctnoprop}.

We also present the dynamical properties of our targets in Table \ref{tab:cctnoprop} alongside their absolute $V$ band magnitudes, $H_{V}$. $H_V$ values for Borasisi-Pabu, 2001~QY$_{297}$, and 2000~OK$_{67}$ were taken directly from the published database of the TNOs are Cool survey \citep{2012A&A...541A..94V,2014A&A...564A..35V}. $H_V$ for 2001~HZ$_{58}$ was calculated using the $H_R$ magnitude and $V-R$ color of this object from \citet{2015A&A...577A..35P}. Because the $H_R$ values published by \citet{2015A&A...577A..35P} are not corrected for phase darkening we must account for the possibility that our value of $H_V$ is overestimated. $\beta_V=0.157\pm0.017~mag/^{\circ}$ was adopted as a nominal phase coefficient by averaging the values of $\beta_V$ reported by \citet{2007AJ....133...26R} for TNOs with $H_V>4$. From its JPL Horizons ephemeris, we determined that 2001~HZ$_{58}$ has never been observed at a phase angle greater than $\alpha=1.35^{\circ}$ since discovery. Based on this maximum phase angle and nominal phase coefficient, we predict that $H_V$ may be overestimated by up to 0.21 magnitudes, hence the asymmetric uncertainty of $H_V$ for 2001~HZ$_{58}$ reported in this work. To estimate $H_V$ for our Col-OSSOS targets, 2013~UL$_{15}$ and 2014~UE$_{225}$, we first used the $H_r$ and $g-r$ values from \citet{2019ApJS..243...12S} to get $H_g$. We then followed the example of \citet{2010AJ....139.1394S,2012AJ....144..169S} and used $V=g-0.55(g-r)-0.03$ from \citet{2002AJ....123.2121S} to determine $H_V$. Finally, we corrected $H_V$ for phase darkening using our nominal $\beta_V$ and the phase angles of these two targets at the time they were observed by Col-OSSOS \citep{2019ApJS..243...12S}.

Thorough statistical analysis of the properties of such a small sample of TNOs would be a foolhardy effort, but we did compare their measured reflectance properties to their dynamical properties and $H_V$ values by eye to see if any potential trends in the data might be glimpsed. No convincing trends were observed, however. Neither do the binary TNOs, Borasisi-Pabu and 2001~QY$_{297}$, exhibit any spectral behaviour that differentiates them from the rest of the objects in our sample, nor do we observe any discernible relation between the reflectance properties of our targets and dynamical residence within, or near to, the cold classical kernel \citep[][]{2011AJ....142..131P,2016AJ....152...70B}.

\section{Interlopers}\label{sec:interlopers}
\citet{2017AJ....154..101P} put forward the possibility that observers may be able to distinguish hot classical TNOs from cold classical TNOs by the shape of their reflectance spectra at $\lambda>0.8~\mu$m. Here we present the first spectroscopic implementation of this technique by reporting that 2000~OK$_{67}$ is very likely to be a hot classical interloper in cold classical orbital parameter space. 2000~OK$_{67}$ has been variably attributed membership in both the hot and cold classical dynamical classes \citep[e.g.][]{2015A&A...577A..35P,2020tnss.book..153M}, largely depending on where researchers have chosen to divide the two populations in terms of inclination. While our assignment of 2000~OK$_{67}$ to the hot classicals is partly based on the fact that its reflectance spectrum does not level out toward the NIR nearly as much as those of the other cold classical targets in our sample, it is important to note that distinction of hot classical TNOs from cold classical TNOs cannot be done by characterisation of their z-band reflectance properties alone. Two of the cold classical TNOs observed by Col-OSSOS have colors which suggest that their reflectance spectra may be close to linear from $0.4-1.0~\mu$m \citep[][]{2017AJ....154..101P}. Indeed, Arrokoth also shares this trait, as its average optical spectral gradient remains approximately constant at least as far to the red as $\sim1.0~\mu$m \citep{Sterneaaw9771,2020Sci...367.3705G}. Our classification of 2000~OK$_{67}$ is based upon the combination of its near linear reflectance spectrum from $0.55-1.0~\mu$m, its relatively low optical redness in comparison to the average for cold classical TNOs (see Table \ref{tab:cctnoprop}), and its already ambiguous dynamical classification. Note that 2000~OK$_{67}$ would not look out of place among the cold classicals when compared in terms of size and albedo \citep[$D=164^{+33}_{-45}$~km and $p_V=0.169^{+0.159}_{-0.052}$ respectively; cf.][]{2020tnss.book..153M}. We caution that multiple properties of a classical TNO must be considered before it may be assigned to either the hot or cold dynamical classes.

\section{Discussion}

\subsection{Silicates}
Based on their finding that the cold classicals occupy a unique region of $g-r$/$r-z$ colour space, \citet{2017AJ....154..101P} inferred that their reflectance spectra have distinct behaviour at $\lambda\sim0.9~\mu$m that might possibly result from the presence of surface material on cold classicals that has an absorption band at those wavelengths, and is not present on the surfaces of the dynamically hot TNOs. We find, however, that there is no obvious absorption band present at $\lambda\sim0.9~\mu$m in any of our reflectance spectra. Inclusion of published $J$-band photometric data for our targets \citep{2003Icar..161..501M,2004A&A...417.1145D,2019ApJS..243...12S} further strengthens this non-detection. Rather than the presence of an absorption band, all three cold classical targets in our sample which have $J$-band photometric measurements exhibit reflectance properties that are more consistent with the existence of featureless linear continuum in their spectra between $\sim1.0~\mu$m and $\sim1.2~\mu$m.

The non-detection of a silicate absorption band is disappointing, but given the low densities so far typically determined for cold classical TNOs it is not surprising \citep[see][]{2014A&A...564A..35V,2020tnss.book..153M}. If anhydrous mafic silicates are present on the surfaces of the cold classicals it is plausible that their strong $1.0~\mu$m absorption bands may be masked by any opaque macromolecular organics they are mixed with \citep[e.g.][]{2008ssbn.book..483D}. Laboratory studies have shown that masking of the $1.0~\mu$m band may be achieved when irradiated organics are present at only half the concentration of the mafic silicate component within a porous refractory mantle \citep[e.g.][]{2016Icar..267..154P}. Because of this possible masking of the $1.0~\mu$m silicate band, our reflectance spectra can neither rule out the presence of silicates on the surfaces of our targets, nor their absence. IR observations of these cold classical TNOs at $\lambda>2.5~\mu$m may fare better in directly detecting the signatures of mafic silicates on the surfaces of cold classical TNOs \citep[e.g.][]{2016PASP..128a8010P}.   

\newpage

\subsection{Carbonaceous Material}
As in previous spectroscopic studies of cold classicals and other extremely red TNOs the reflectance properties of our targets appear most consistent with those of irradiated residues comprised of complex macromolecular organic compounds \citep[e.g.][]{1998Icar..135..389C,2006A&A...455..725B,2006ApJ...644..646B,2020Sci...367.3705G}. Like our targets, such materials have optical reflectance spectra that broadly consist of a linear continuum with a strong positive gradient that may curve downward close to the boundary between the optical and NIR ranges \citep[e.g.][]{2008ssbn.book..483D}.

Delocalized $\pi$-bonded electrons within organic molecules absorb light via $\pi-\pi^{*}$ and $n-\pi^{*}$ excitation; the energy structure of these delocalised electrons is governed by the size and clustering of the conjugated $\pi$-bond networks in which they reside. In turn these factors are influenced by the interconnected compositional and structural properties of the organic molecules themselves, such as the sp$^{2}$/sp$^{3}$ bond ratio, the extent of sp$^{2}$ bond clustering, the abundance ratios of various constituent elements (e.g. C/N, C/H, C/O), and the number of nitrogen and oxygen bearing heterocycles and auxochrome functional groups. In concert all of these factors determine the albedo of an irradiated organic sample or residue, and the gradient and curvature of its reflectance spectrum from the near-UV to the nearest IR \citep[e.g.][]{1996P&SS...44..741M,2004Icar..168..344I,2005AdSpR..36..178C,2006Icar..185..301B,2006ApJ...644..646B,2008ssbn.book..483D}. 

Unfortunately the mapping between the optical reflectance properties of an irradiated organic sample and the properties of its diverse constituent macromolecules is often intractably degenerate \citep{2006Icar..185..301B}. As a result it is not possible to use our reflectance spectra to diagnostically characterise any carbonaceous material present on the surfaces of our targets. Sufficiently sensitive observations at $\lambda>1.0~\mu$m stand a better chance of being useful in efforts to characterise any refractory organics on the surfaces of cold classical TNOs, through detection and characterisation of the vibrational molecular absorption bands that refractory organics often exhibit at IR wavelengths \citep[e.g.][]{2004Icar..168..344I,2008ssbn.book..483D,2014ApJ...788..111M,2015ApJ...812..150M}. Sufficiently sensitive observations of cold classical TNOs at near-UV wavelengths may also be informative about the carbonaceous materials on their surfaces, as carbon-rich phases exhibit multiple near-UV spectral behaviours that are diagnostic of both their composition and physical state \citep{2016M&PS...51..105H,2018Icar..307...40A}.

\subsection{The Cold Classical Surface Type}

While optical reflectance spectra of organics cannot be used for precise characterisation, it is possible to use them to tell different classes of carbonaceous material apart \citep[e.g.][]{2008ssbn.book..483D,2012ApJ...749...33F}. Because organic compounds are predicted to dominate the optical and very-NIR reflectance properties of very red TNOs \citep[e.g.][]{2015Icar..252..311D}, the distinctly consistent overall shape of the reflectance spectra of all our 100~km scale cold classical targets suggests that the class of refractory organics on their surfaces is similar from object to object. If this homogeneity is a property shared by all larger members of the cold classical population, as suggested by their unimodal and relatively narrow color and albedo distributions \citep[][]{2017AJ....154..101P,2019ApJS..243...12S,2020tnss.book..153M}, it supports the prediction that the cold classical TNOs are likely to have both formed under similar conditions and experienced a common history of surface processing. Laboratory studies support this idea by repeatedly showing that if refractory organics are formed from different initial ice mixtures, processed in different ways, or processed to different extents, they will typically exhibit reflectance spectra with large differences in shape, unlike those of the cold classicals \citep[][]{2004Icar..168..344I,2005AdSpR..36..178C,2006Icar..185..301B,2008ssbn.book..483D,2014ApJ...788..111M,2015ApJ...812..150M,2018ApJ...856..124P}. 

While the composition and molecular structure of any refractory organics on the surfaces of our targets may define their reflectance properties, the role of the physical properties of their surfaces, such as grain size distribution or porosity, must not be ignored. The physical properties of the surface of a TNO may determine the properties of its reflectance spectrum independently of composition \citep[e.g.][]{2016Icar..267..154P,2018Icar..305..203C}. The homogenous shape of the reflectance spectra of cold classical TNOs may therefore indicate that their surfaces have similar physical properties, while the physical properties of the surfaces of the dynamically hot TNOs may be more diverse. Unfortunately, without high phase angle observations of TNOs it is not currently possible to disentangle any effects that the physical properties and composition of their surfaces have on their reflectance spectra. High phase angle measurements from spacecraft will be required to examine the physical properties of TNO surfaces \citep[see][]{2016ApJ...828L..15P}.

It is worth noting that some of the reflectance spectra of dynamically hot TNOs (here in which we include the centaurs), also exhibit a flattening of their optical gradients toward the NIR \citep[e.g.][]{2008A&A...487..741A,2009A&A...508..457F,2010Icar..208..945M}. In particular the behaviour appears more commonly in the reflectance spectra of TNOs with similar optical redness to the cold classicals. In contrast to that of the cold classicals, however, the flattening behaviour observed for dynamically hot TNOs is diverse in both shape and the wavelength at which it occurs. Reports show that the spectra of dynamically hot TNOs can begin to diverge from their linear continuum slopes anywhere within the range $0.65-1.4~\mu$m \citep[e.g.][]{1998Icar..135..389C,2004A&A...415L..21B,2006A&A...455..725B,2008A&A...487..741A,2004A&A...421..353F,2009A&A...508..457F,2015A&A...582A..13G,2005A&A...444..977M,2010Icar..208..945M,2017A&A...604A..86M,2019ApJS..243...12S}. By comparison, all of our cold classical reflectance spectra flatten within the narrower range of $0.75-0.9~\mu$m. Therefore, while flattening behaviour is not a unique quality of the reflectance spectra of cold classical TNOs, the shape of their reflectance spectra is remarkably consistent in comparison to those of the dynamically hot TNOs. Our cold classical reflectance spectra support the prediction of \citet{2017AJ....154..101P}, that the cold classical TNOs (at least those observable from current ground-based facilities) comprise a unique surface type among TNOs. Not only do the cold classicals have distinct distributions in terms of optical color and albedo, they also appear to have reflectance spectra with a consistent characteristic shape from optical wavelengths to the nearest IR.     

\subsection{Comparison to Arrokoth}\label{sec:arrokoth}

\citet{2019Icar..334...22B} noted that their photometric measurements of cold classical TNOs hinted at the existence of an interesting, but not statistically significant, increase in the $F606W-F814W$ color of cold classicals of decreasing size. The F814W filter used by \citet{2019Icar..334...22B} has a bandpass covering $0.7-0.97~\mu$m, and includes the region in which our cold classical reflectance spectra flatten. An increase in $F606W-F814W$ color for smaller cold classicals may therefore be interpreted as a shift toward longer wavelengths of the point at which a reduction in the spectral gradient occurs.

We note that not one of the five, nearly randomly selected, $\sim100$~km scale cold classical TNOs in our sample has a reflectance spectrum like that of the $\sim10$~km scale Arrokoth, which has linearly increasing reflectance at least as far to the red as $\sim1.0~\mu$m. The potential size-color trend hinted at by \citet{2019Icar..334...22B}, and the absence of linear spectral behaviour in our sample at $\lambda>0.8~\mu$m raises the question as to whether spectral flattening at $\lambda>0.8~\mu$m is more common in the reflectance spectra of large cold classicals than it is in those of small ones.

Our spectra unfortunately do not have sufficient NIR wavelength coverage for us to precisely measure the central wavelength of their downward curvature at $\lambda>0.8~\mu$m. Nor is our sample large enough for statistically significant analysis. As a result, further speculation on this suggested trend without a stronger foundational dataset would be reckless. We finish here by stating that the size-color trend for cold classical TNOs hinted at by \citet{2019Icar..334...22B} appears, at least in broad qualitative terms, to be consistent with the absence of linear spectral behaviour at $\lambda>0.8~\mu$m in the reflectance spectra of all our $\sim100~$km scale cold classical targets.  

\section{Conclusions}
Following reports by \citet{2017AJ....154..101P} and \citet{2018PASJ...70S..40T} that the cold classical TNOs have distinct photometric color properties, and potentially even an absorption band in their reflectance spectra at $\lambda>0.8~\mu$m, we observed the reflectance spectra of six TNOs currently residing in the cold classical dynamical range. Our reflectance spectra, which cover $0.55-1.0~\mu$m, are the first to be observed in the optical range for any of our targets. We find that the reflectance spectra obtained are consistent with coarse reflectance spectra derived from previously published photometric colors for our targets. 

Five of the six targets we observed, including Borasisi-Pabu, 2001~QY$_{297}$, 2001~HZ$_{58}$, 2013~UL$_{15}$, and 2014~UE$_{225}$, have reflectance spectra which are linear and red in the range $0.55<\lambda<0.75~\mu$m, but show a broadly consistent flattening at $\lambda>0.8~\mu$m. This result is consistent with the reported lower average $r-z$ color of cold classicals in comparison to dynamically hot TNOs with similar $g-r$ color, and the suggestion that the cold classicals occupy a distinct surface class \citep{2017AJ....154..101P}. We find no evidence that the observed flattening in our spectra is associated with the presence of an absorption band at $\lambda\sim1.0~\mu$m; anhydrous mafic silicates remain elusive on the surfaces of TNOs. We predict that the similar shape of the reflectance spectra of these five targets arises due to the presence of similar refractory organics on their surfaces and/or similarity between the physical properties of their surfaces. Such apparent consistency between the reflectance properties of cold classicals aligns well with their unimodal and relatively narrow color and albedo distributions, supporting predictions that the cold classical TNOs formed under similar conditions and have since experienced similar types and extents of surface evolution. 

We assert that the sixth TNO in our sample, 2000~OK$_{67}$, as likely to be a hot classical TNO interloping in the cold classical dynamical range. While this is partly because the reflectance spectrum of 2000~OK$_{67}$ is close to linear across the full $0.55-1.0~\mu$m range, we caution that the presence or absence of flattening at $\lambda>0.8~\mu$m in a TNO's reflectance spectrum is insufficient on it's own for the purposes of classification. Our assignment of 2000~OK$_{67}$ to the hot classicals is also based upon its sporadic assigment to the hot classicals in prior reports, and its relatively low optical spectral gradient in comparison to the average for cold classical TNOs.

Future sufficiently sensitive observations at near-UV and IR wavelengths will be crucial in making headway in determining whether refractory organics are responsible for the distinct reflectance properties of cold classical TNOs, as these wavelength regions have been shown to exhibit features that are diagnostic of both the chemical composition and molecular structure of complex carbonaceous materials \citep[e.g.][]{2014ApJ...788..111M,2015ApJ...812..150M,2016M&PS...51..105H,2018Icar..307...40A}. Observations of cold classical TNOs at high phase angles \citep[which are only feasible from vantage points at large heliocentric distances; e.g.][]{2016ApJ...828L..15P} will also be required to explore any possible connection between the physical properties and the distinct reflectance properties of cold classical surfaces.

\acknowledgments

We are grateful to Jonathan Smoker, Cedric Ledoux, Joe Anderson, and Francisco Belmar for sharing their expertise and assisting our VLT observations. Thanks also to Xiaoyu Zhang and Dale Cruikshank for their assistance in accessing literature on refractory organics. T.S. is supported through a Gemini Science Fellowship by the international Gemini Observatory, a program of NSF’s OIR Lab, which is managed by the Association of Universities for Research in Astronomy (AURA) under a cooperative agreement with the National Science Foundation, on behalf of the Gemini partnership of Argentina, Brazil, Canada, Chile, the Republic of Korea, and the United States of America. T.S. was also supported in part by the Astrophysics Research Centre at Queen's University Belfast, and the Northern Ireland Dept.~for the Economy. This work is based on observations collected at the European Organisation for Astronomical Research in the Southern Hemisphere under ESO program 0103.C-0708. This research made use of NASA’s Astrophysics Data System Bibliographic Services, the JPL HORIZONS web interface (https://ssd.jpl.nasa.gov/horizons.cgi), and data and services provided by the IAU Minor Planet Center.

%

\vspace{5mm}
\facility{ESO: VLT-UT1(FORS2)}


\software{Astropy \citep{2013A&A...558A..33A}, ESO Reflex \citep{2013A&A...559A..96F}, Matplotlib \citep{2007CSE..9..90H}, NumPy \citep{harris2020array}, SciPy \citep{2020SciPy-NMeth}}



\bibliography{masterbib}{}
\bibliographystyle{aasjournal}

\appendix

\begin{figure}[b!]
\centering
\includegraphics[scale=0.8]{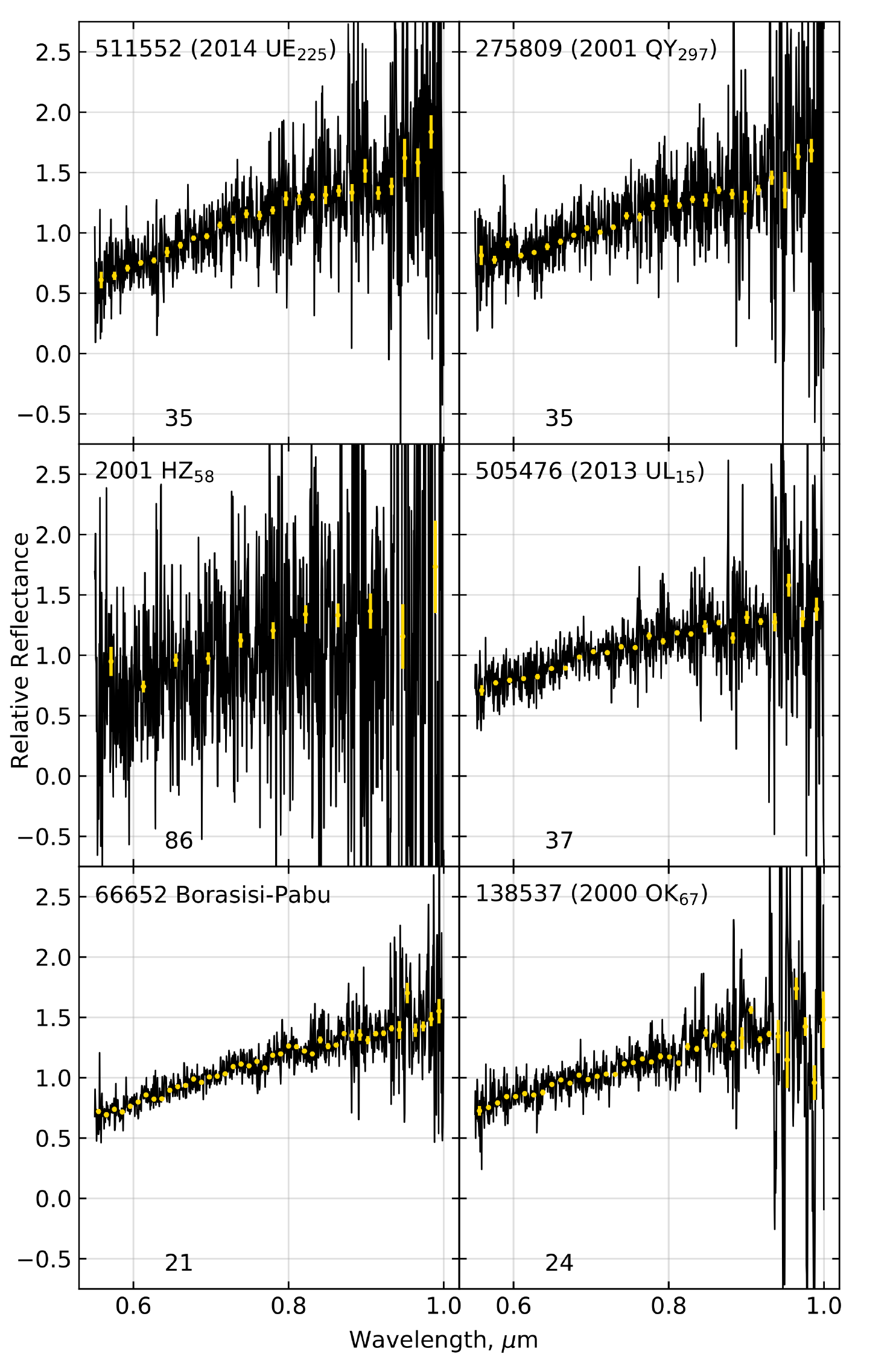}
\caption{The reflectance spectra presented in Fig. \ref{fig:spec}, but prior to spectral binning, are plotted here in black. The associated binned spectra are plotted with yellow points. The binning factor (i.e. number of resolution elements per bin) used to produce the binned spectrum from the unbinned one is presented at the bottom of each panel.  \label{fig:unbin}}
\end{figure}

\begin{figure}
\centering
\includegraphics[scale=0.8]{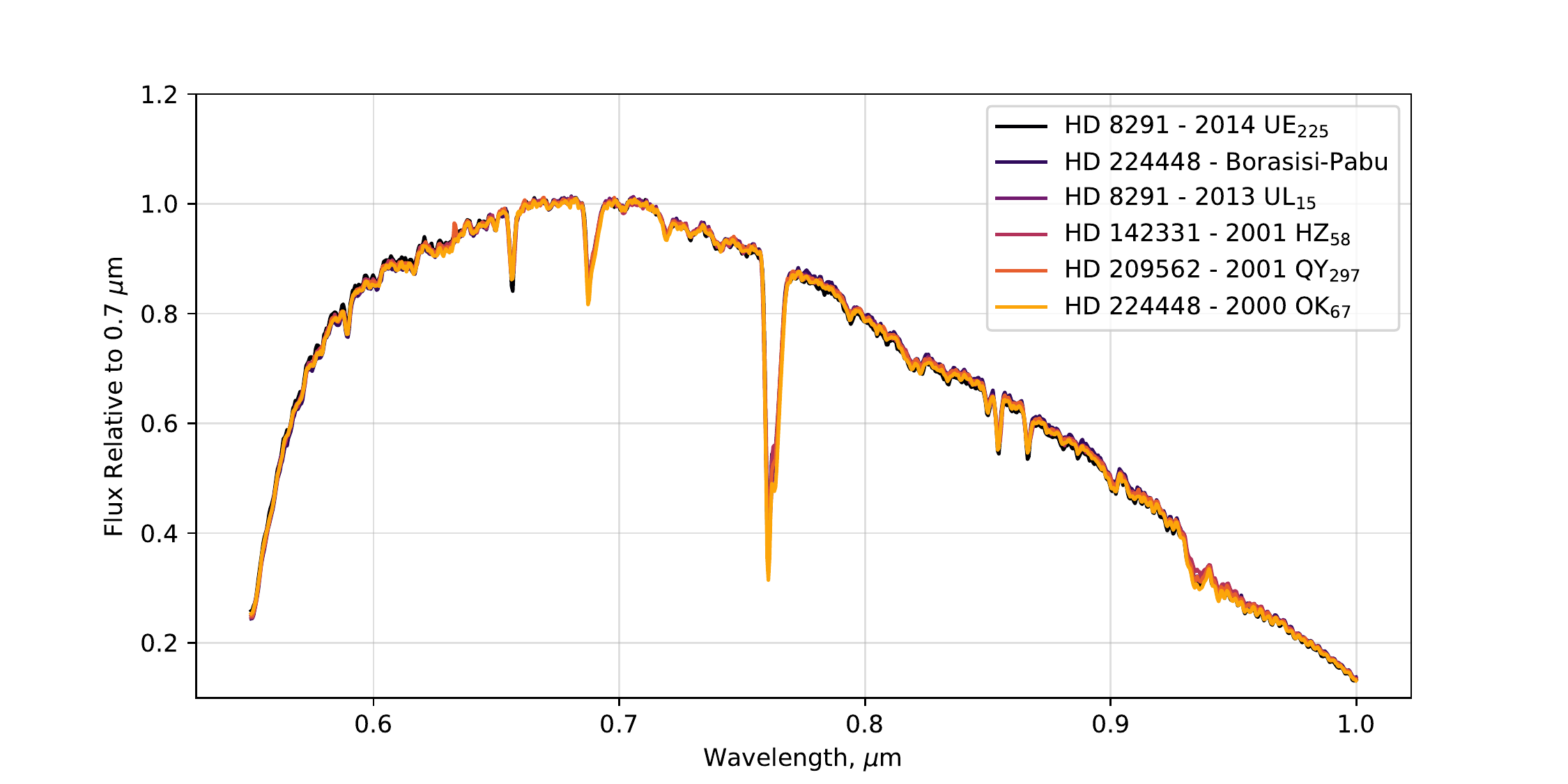}
\caption{The unbinned stacked spectra of the Solar twins used to calibrate each of our TNO spectra. They are scaled to unity at $0.7~\mu$m, but are not corrected for instrument response. The shapes of the star spectra are very consistent across our full wavelength coverage, with the only noteworthy variation between them occuring in bands of moderate telluric absorption at $\lambda\sim0.77~\mu$m and $0.93-0.95~\mu$m. The consistency of our calibrator star spectra shows that any variation between the reflectance spectra of our TNO targets primarily results from differences between the intrinsic reflectance properties of the TNOs themselves. \label{fig:stars}}
\end{figure}

\begin{figure}
\centering
\includegraphics[scale=0.6]{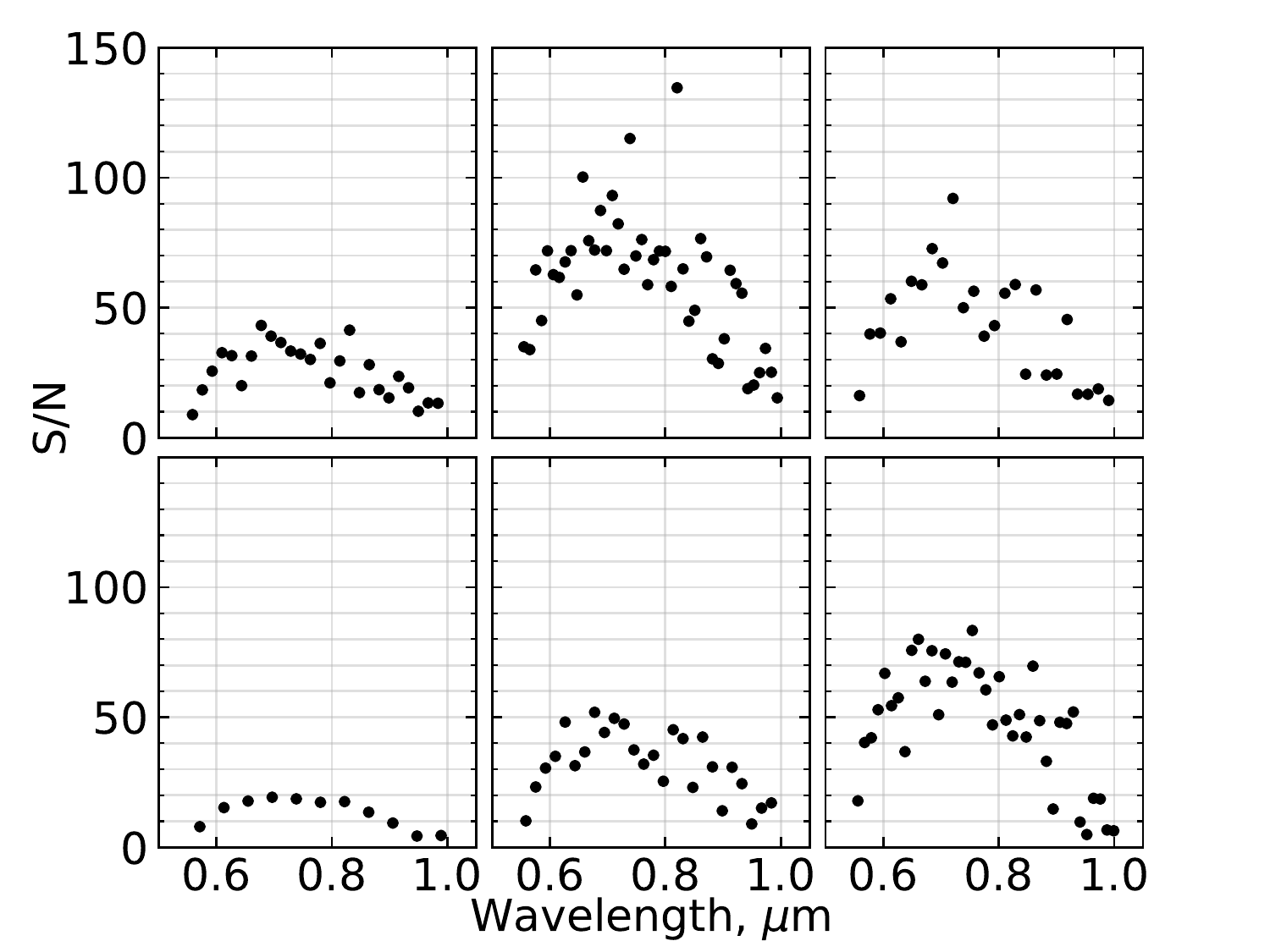}
\caption{The wavelength dependent S/N of our reflectance spectra. Going left to right along the top row and then the bottom row, the panels present data for 2014~UE$_{225}$, Borasisi-Pabu, 2013~UL$_{15}$, 2001~HZ$_{58}$, 2001~QY$_{297}$, and 2000~OK$_{67}$. \label{fig:snr}}
\end{figure}



\end{document}